\documentclass[12pt]{iopart}
\usepackage{cite}
\usepackage[dvips]{graphicx}

\begin{document}

\title{The effect of bias feed profile on spectral properties of noisy Josephson flux flow oscillator}

\author{E A Matrozova$^{1,2}$, A L Pankratov$^{2,1}$, L S Revin$^{2,1}$}
\address{$^1$Laboratory of Cryogenic Nanoelectronics, Nizhny Novgorod State Technical University, Nizhny Novgorod, Russia
\\$^2$Institute for Physics of Microstructures of RAS, GSP-105, Nizhny Novgorod, 603950, Russia}
\ead{alp@ipm.sci-nnov.ru}

\begin{abstract}
For creation of a noisy non-stationary spectrometer working in the subTHz frequency range the power and spectral characteristics of a flux flow oscillator (FFO), based on a long Josephson junction are studied. The effect of various bias current profiles on the spectral linewidth of long overlap Josephson junction are investigated and comparison with the inline junction geometry is outlined. It has been demonstrated that the spectral linewidth can be increased by a factor of three with the moderate reduction of emitted power.
\end{abstract}

\maketitle
\nosections

At the present time for many applications including ecology and atmosphere monitoring, hi-tech and safety systems, medicine and biology, the problem of the precise analysis of multicomponent gas composites is important. For a wide variety of problems, it is necessary to perform the detection of microconcentrations of several gases simultaneously that imposes the strong requirement of sensitivity and speed of a spectrometer operation. New method of nonstationary microwave spectroscopy based on noise sources of radiation allows to realize high sensitivity, high resolution and possibility of wide range measurements simultaneously\cite{vks}. The Josephson flux flow oscillator\cite{ks} (FFO) based on a long Nb-AlOx-Nb Josephson junction, generating the broadband signal, is a good candidate to be a radiation source for creation of a noisy spectrometer. Besides this, it is very important to have a good noisy source for calibration of mixers, such as based on SIS (superconductor-insulator-superconductor) Josephson junctions, and for calibration of superconductive integrated receiver (SIR) \cite{sir,sir2}.

The FFO \cite{ks} possesses a number of benefits in comparison with other oscillators of microwave range, such as compactness, small power consumption and possibility to work in subTHz and THz ranges of frequencies where the most intensive lines of absorption of many substances lie.
Since the use of a quasi-chaotic oscillation regime of the FFO \cite{ukh} seems to be not suitable for practical applications, first of all due to expected small radiation power \cite{mplv},
it is very important to achieve large enough spectral linewidth and radiation power in the range of 450-700 GHz, where continuous frequency tuning is possible. Besides, it has been demonstrated recently that if the power spectrum of a noisy signal has the Lorentzian shape, then the spectroscopy can be performed in a similar manner as for coherent signals \cite{spv}. While the FFO has perfectly Lorentzian spectral line \cite{KSsust}, its linewidth is restricted by 1-50 MHz in 450-700 GHz working range, which is not enough for variety of applications. The aim of the present paper is to study the possibility to increase the FFO linewidth without serious degradation of its radiation power, by variation of biasing electrodes configuration.

It is known that all basic properties of the FFO can be described in the frame of the sine-Gordon equation:
\begin{equation}
{\phi}_{tt}+\alpha{\phi}_{t}
-{\phi}_{xx}=\beta{\phi}_{xxt}+\eta(x)-\sin (\phi)+\eta_f(x,t)
\end{equation}
where indices $t$ and $x$ denote temporal and spatial derivatives, $\phi$ is the phase order parameter. Space and time are normalized to the Josephson penetration length
$\lambda _{J}$ and to the inverse plasma frequency
$\omega_{p}^{-1}$, respectively, $\alpha={\omega_{p}}/{\omega_{c}}$
is the damping parameter, $\omega_p=\sqrt{2eI_c/\hbar C}$,
$\omega_{c}=2eI_cR_{N}/\hbar$, $I_c$ is the critical current, $C$ is
the Josephson tunnel junction (JTJ) capacitance, $R_N$ is the normal state resistance, $\beta$
is the surface loss parameter, $\eta$ is the dc overlap bias
current density, normalized to the critical current density $J_c$,
and $\eta_f(x,t)$ is the fluctuational current density. If the
critical current density is fixed and the fluctuations are treated
as white Gaussian noise with zero mean, its correlation function is:
$\left<\eta_f(x,t)\eta_f(x',t')\right>=2\alpha\gamma \delta
(x-x^{\prime})\delta (t-t^{\prime})$, where $\gamma = I_{T} /
(J_{c}\lambda_J)$ is the dimensionless noise intensity,
$I_{T}=2ekT/\hbar$ is the thermal current, $e$ is the electron
charge, $\hbar$ is the Planck constant, $k$ is the Boltzmann
constant and $T$ is the temperature.

The boundary conditions, that simulate simple RC-loads, see Ref.s
\cite{Parment,pskm,p,pa}, have the form:
\begin{eqnarray}\label{x=0}
\phi(0,t)_{x}+r_L c_L\phi(0,t)_{xt}-c_L\phi(0,t)_{t t}+\\
\beta r_L c_L\phi(0,t)_{xtt}+\beta\phi(0,t)_{x
t}=\Gamma, \nonumber \\ \phi(L,t)_{x}+r_R
c_R\phi(L,t)_{x t}+c_R\phi(L,t)_{t t}+ \label{x=L} \\  \beta r_R
c_R\phi(L,t)_{xtt}+\beta\phi(L,t)_{x t}=\Gamma.
\nonumber
\end{eqnarray}
Here $\Gamma$ is the normalized magnetic field, and $L$ is the
dimensionless length of the FFO in units of $\lambda _{J}$. The dimensionless capacitances and
resistances, $c_{L,R}$ and $r_{L,R}$, are the FFO RC-load placed at
the left (output) and at the right (input) ends, respectively.

In general, there are two possible ways to increase the radiation linewidth of the FFO at flux flow steps (i.e. in the frequency range 450-700 GHz). The simplest way is mixing either the bias current or the magnetic field with low-frequency noise. However, the Lorentzian shape of spectral line then will be transformed into the Gaussian one. Another way is to adjust the bias current profile in an optimal fashion. Namely this second way is the subject of the present paper.

\begin{figure}[ht]
\resizebox{1\columnwidth}{!}{
\includegraphics{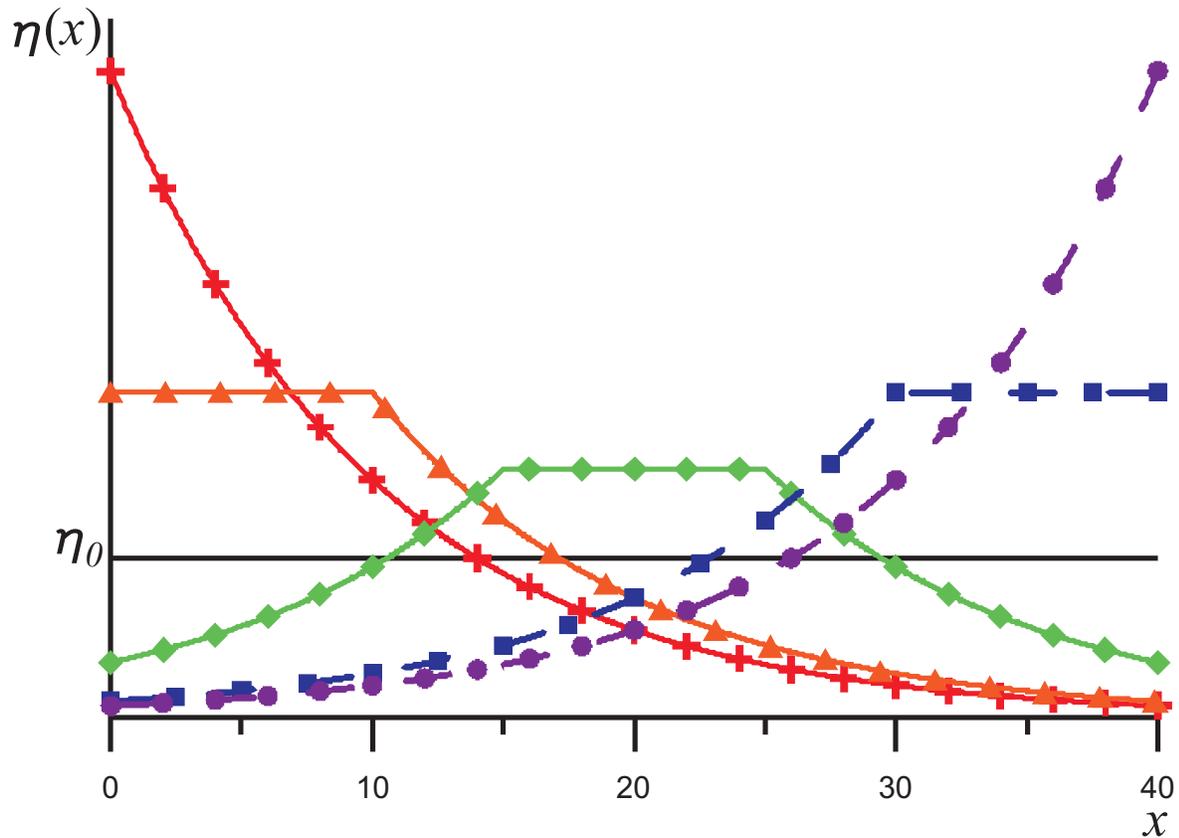}}
{\caption{Bias current profiles: thin solid line - uniform distribution, curve with crosses corresponds to the extremely narrow electrode shifted to the left ($x_{1}=x_{2}=0$), curve with triangles - ($x_{1}=0$, $x_{2}=10$), curve with diamonds - electrod position at the center of the junction ($x_{1}=15$, $x_{2}=25$), curve with squares - ($x_{1}=30$, $x_{2}=40$), curve with cirlces - ($x_{1}=x_{2}=40$).}
\label{ix1}}
\end{figure}
The spatial distribution $\eta(x)$ of the bias current is determined by the configuration of the biasing electrodes used for current supply. The simplest case of uniform distribution (Fig. \ref{ix1}, thin solid line) is achieved by the equality of the biasing electrode length to the junction length. However, this is an ideal case: obtaining a uniform profile in practice requires additional tricks, such as a normal metal insertions since the current in the superconducting electrode increases with approaching the boundaries of the electrode. If the electrode is narrower than the Josephson junction, the current decreases away from the electrode edge (Fig. \ref{ix1}, curves with symbols). In this case one says that the JTJ has the so-called "unbiased tail". The position of the biasing electrode is described by two points $x_1$ and $x_2$ $(0 \leq x_1 < x_2 \leq L)$, which defines the left and the right boundaries, respectively. Current profile can be divided into three parts. The central part corresponding to the current along the electrode is accepted for simplicity to be uniform $(x_1 \leq x \leq x_2)$. In the unbiased regions $(0 \leq x < x_1, x_2< x \leq L)$ the decay of the current profile is assumed to be exponential with $\exp(-px)$ law, where $p$ is the decay factor. If the current electrode is extremely narrow and shifted to one of the junction edges, the current profile has a decaying part only (Fig. \ref{ix1}, curve with crosses and curve with circles).

The computer simulations are performed with the following parameters: junction length $L= 40$, external magnetic field $\Gamma= 3.6$, parameters of losses $\alpha= 0.033$ and $\beta= 0.035$, noise intensity $\gamma= 0.05$, $c_{L}=c_{R}= 100$, $r_{L}= 2$, $r_{R}= 100$, decay factor $p= 0.1$.

First, let us consider a situation where the width of the current electrode is smaller than the length of the Josephson junction. Let us move the electrode along the junction from position with $x_{1}= 0$ to position with $x_{2}= L$. Depending on the electrode position the obtained current profile has one of the forms shown in Fig. \ref{ix1}. Bias current profiles, shifted from the center of the junction to the input (right) edge are shown as dashed.

\begin{figure}[ht]
\resizebox{1\columnwidth}{!}{
\includegraphics{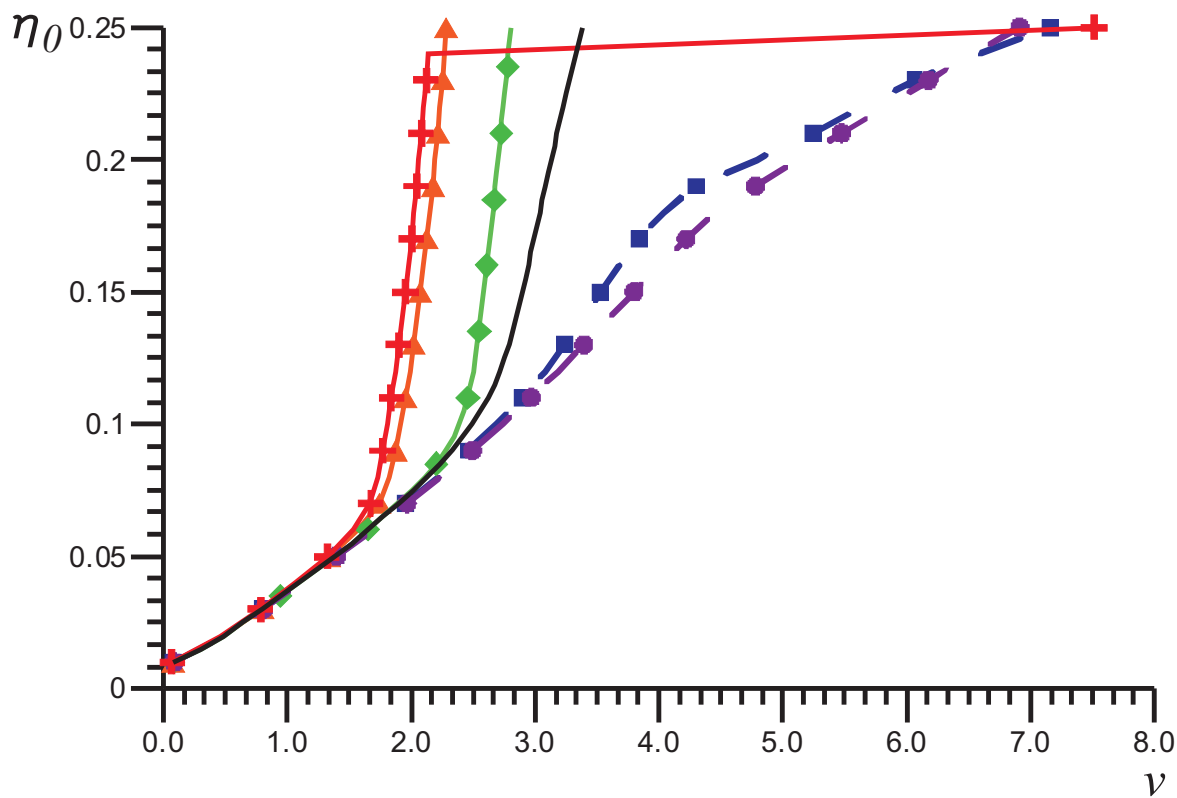}}
{\caption{IV characteristics of the FFO at various bias current profiles, depicted in Fig. \ref{ix1}.}
\label{cv1}}
\end{figure}
The current-voltage curves (IVC) are shown in Fig. \ref{cv1} (the notations are the same as in Fig. \ref{ix1}). The farther from the radiating junction edge electrode is moved, the less steep the IV characteristic is.

\begin{figure}[ht]
\resizebox{1\columnwidth}{!}{
\includegraphics{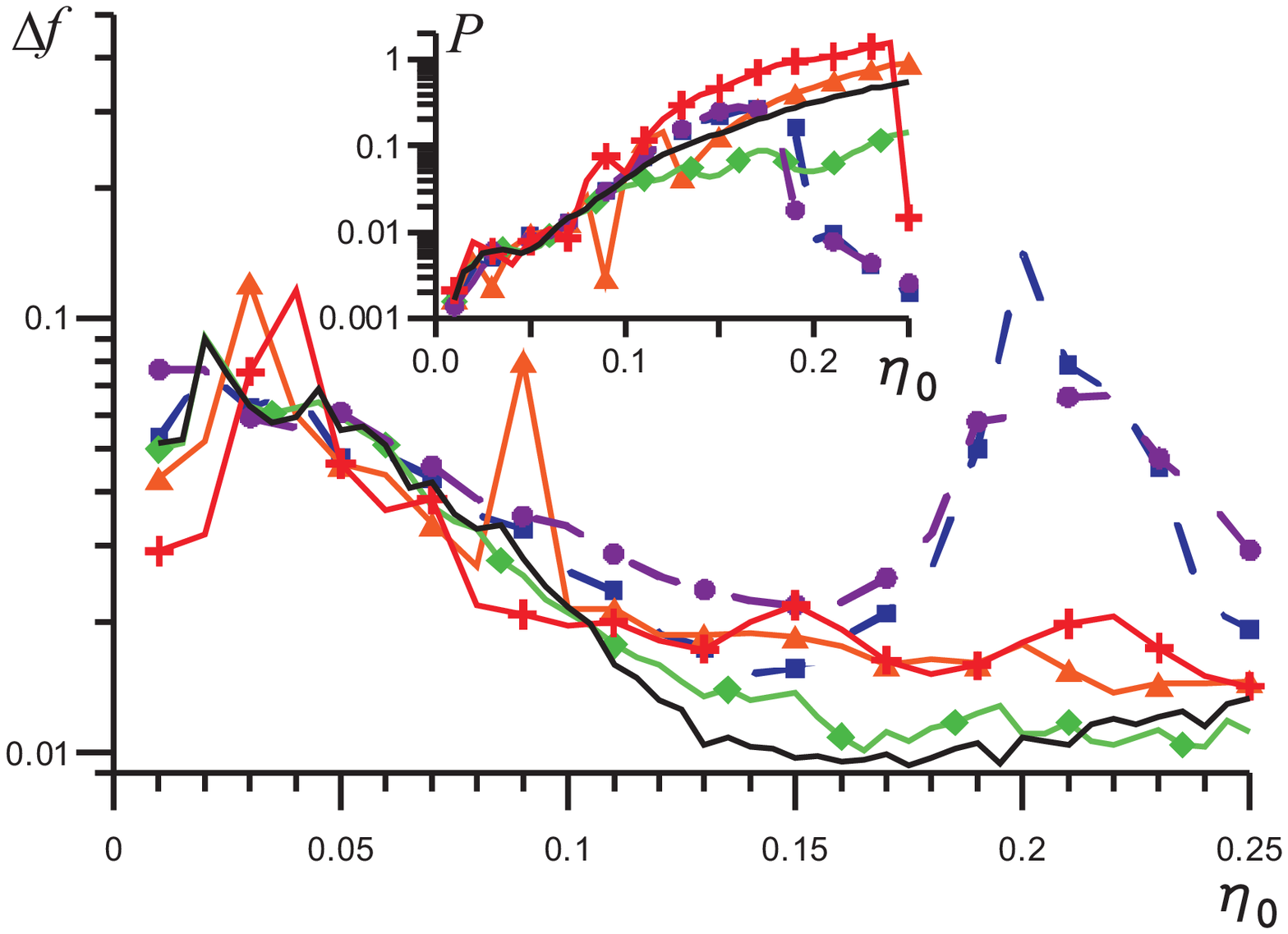}}
{\caption{FFO linewidth for different current profiles, see Fig. \ref{ix1}. Inset: Radiated power for various current profiles, the notations are the same as for the linewidth.}
\label{df1}}
\end{figure}
The spectral linewidths for different current profiles from Fig. \ref{ix1} are presented in Fig. \ref{df1}. As one can see from Fig. \ref{df1}, for total bias current $\eta_0 \sim 0.12 \div 0.22$ the linewidth of a nonuniform current profiles is significantly larger than the linewidth of a uniform profile, while for shorter unbiased tail one can get a reduction of the linewidth, see Ref. \cite{p}. When the current electrode is located near the left edge of the junction ($0 \leq x_1 < x_2 \leq  L/2$), it can get a line broadening factor of about 2 (Fig. \ref{df1}, curve with crosses and curve with triangles). When the electrode is near the right edge ($L/2\leq x_1 < x_2 \leq L$), the broadening of the linewidths is even larger (Fig. \ref{df1}, dashed curve with squares and dashed curve with circles). The sharp increasing of the line in 5-7 times for these curves ($\eta_0 \sim 0.18 \div 0.22$) corresponds to the area of low power (see the inset of Fig. \ref{df1}) and has no interest. Thus, the farther from the radiating edge of the junction electrode is moved, and, consequently, the more flat the voltage-current characteristic is, the broader the spectral line of radiation is obtained (in accordance with the behavior of differential resistance).

The FFO radiated power is presented in the inset of Fig. \ref{df1}. It is seen that the larger the current fed to the output edge of the junction, the greater the power radiated from it. Maximum power is achieved in the case of a narrow electrode shifted to the left edge of the junction. A high power level for a uniform profile of the bias current can also be noted (Fig. \ref{df1}, thin solid curve). The electrode shifted to the right edge of the junction has an obvious maximum of power and for total bias current $\eta_0 \sim 0.15$ the output radiated power is close to the maximum (dashed curves with squares and circles).

The simulation of the shift of the narrow current electrode along the Josephson junction has led to the following conclusions. To obtain a broad spectral line the bias current profile should be nonuniform, i.e., the electrode must be more narrow than the junction. The best result is achieved when the electrode is near the nonradiating junction edge. In addition, this condition allows to receive an acceptable level of radiated power and it is larger than the level for the case of electrode located at the center of the JTJ \cite{pskm}.

\begin{figure}[ht]
\resizebox{1\columnwidth}{!}{
\includegraphics{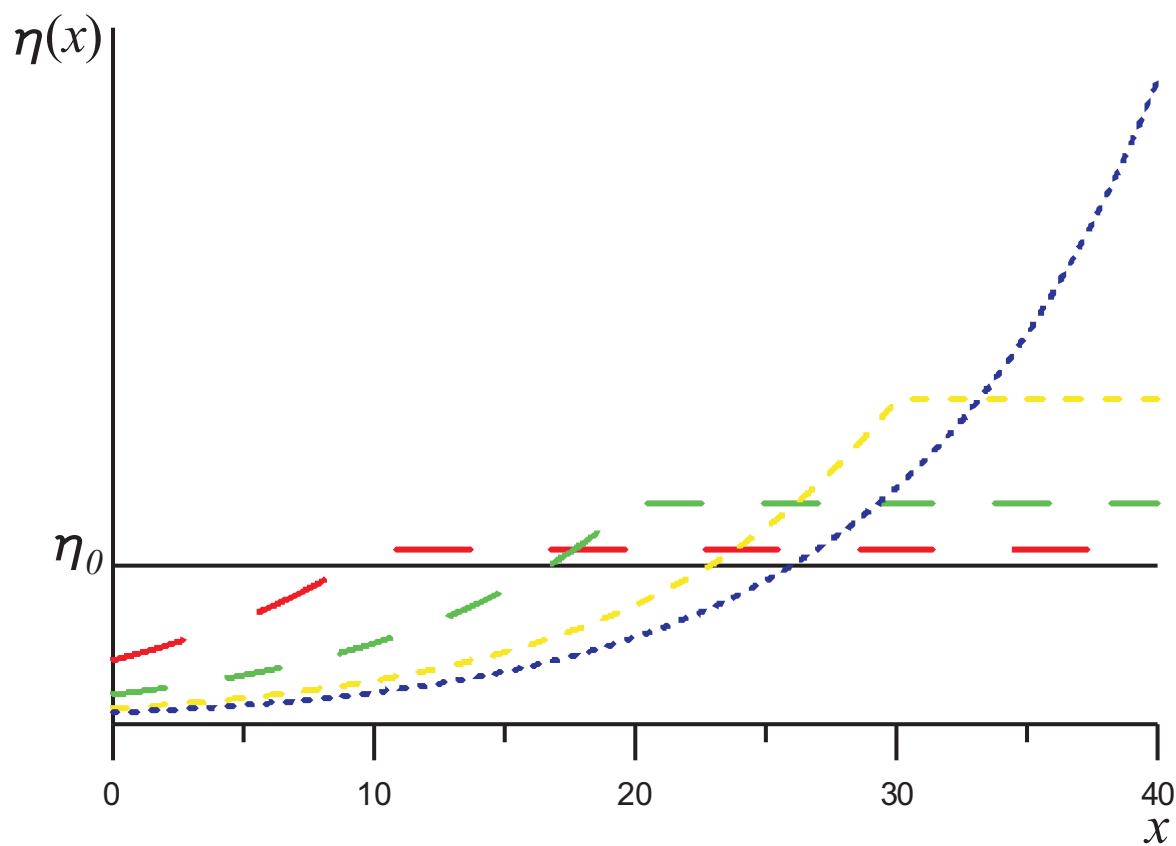}}
{\caption{Bias current profiles: thin solid line - uniform distribution, long-dashed curve -  $x_{1}=10$. dashed curve - $x_{1}=20$, short-dashed curve - $x_{1}=30$, dotted curve - $x_{1}=40$.}
\label{ix2}}
\end{figure}
\begin{figure}[ht]
\resizebox{1\columnwidth}{!}{
\includegraphics{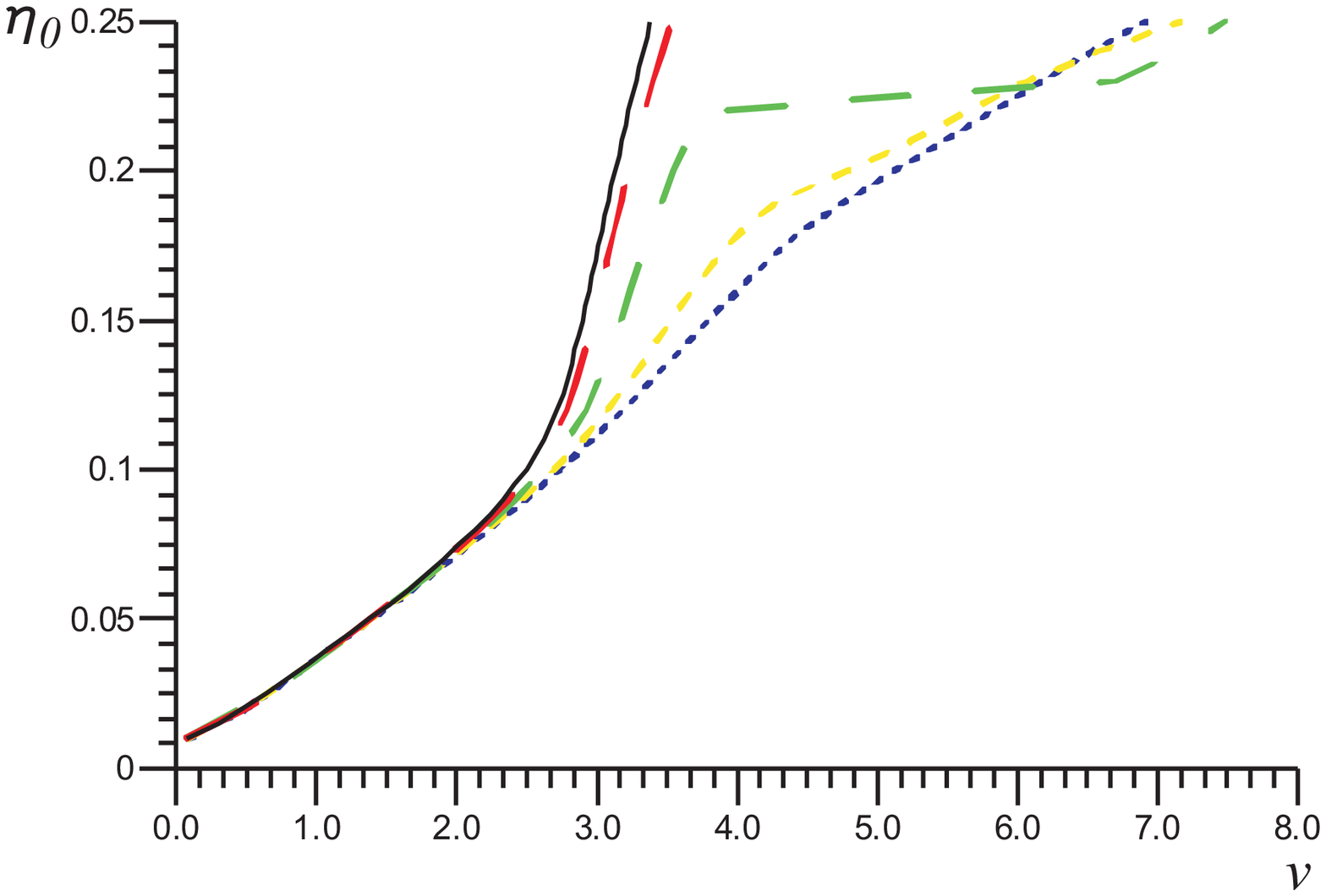}}
{\caption{IV characteristics of the FFO at various bias current profiles presented in Fig. \ref{ix2}}
\label{CV2}}
\end{figure}

Taking into account the above obtained results, let us consider the dependence of the junction characteristics on the width of the electrode when it is shifted to the right edge ($x_{2}=L$). Varying the position of the left boundary of the electrode $x_{1}$, the following current profiles shown in Fig. \ref{ix2} can be obtained. The IVCs are shown in Fig. \ref{CV2} (the notations are the same as in Fig. \ref{ix2}). As can be seen from the figure, a narrower current electrode gives a more flat IVC, and in the limit of the electrode width tending to zero ($x_{1}=x_{2}=40$) the IV characteristic tends to the ohmic line (Fig. \ref{CV2}, dotted curve). The IVC of the uniform current has the largest slope (Fig. \ref{CV2}, thin solid curve).

\begin{figure}[ht]
\resizebox{1\columnwidth}{!}{
\includegraphics{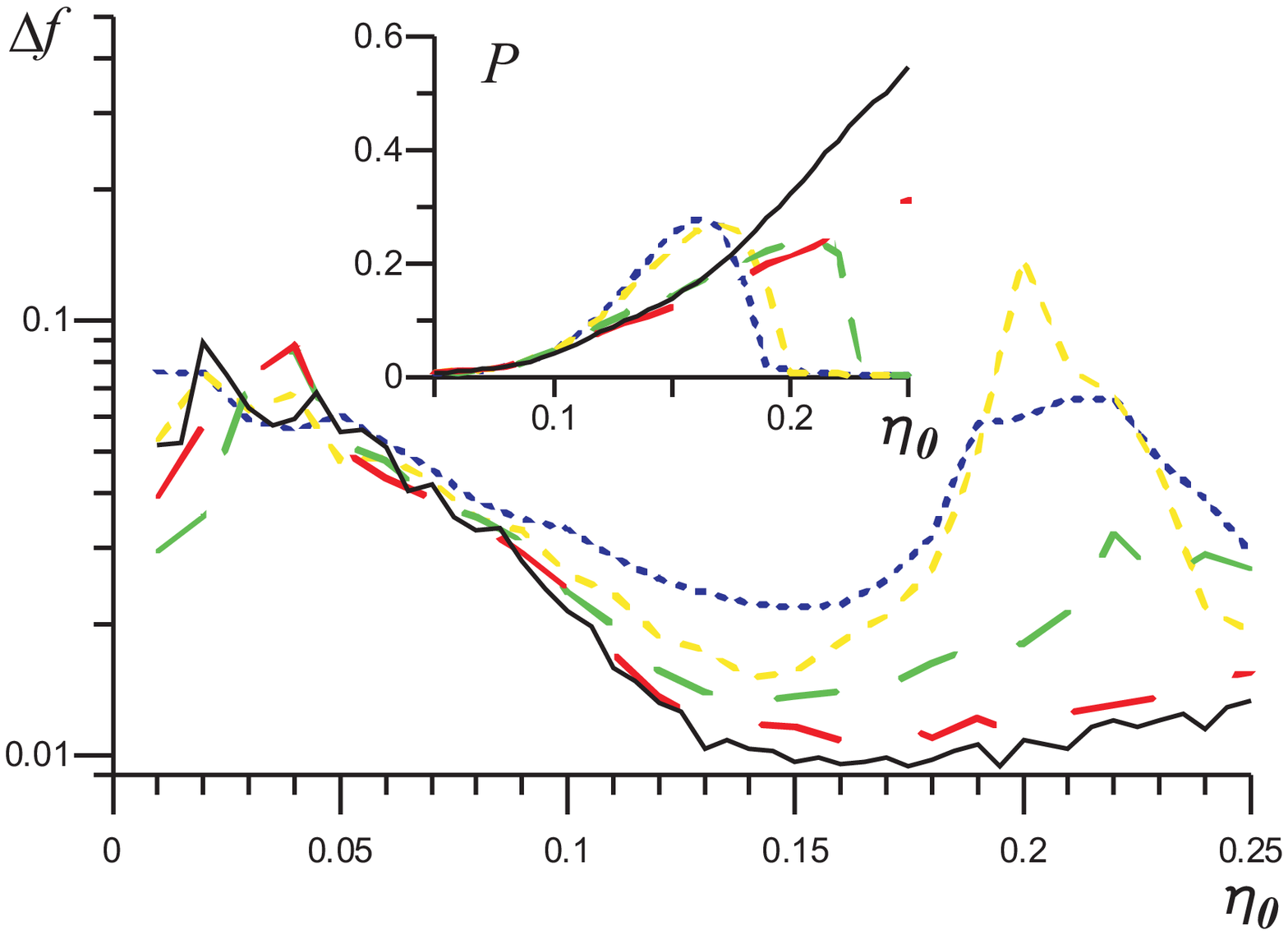}}
{\caption{FFO linewidth for different current profiles, see Fig. \ref{ix2}. Inset: Radiated power for various current profiles, the notations are the same as for the linewidth.}
\label{df2}}
\end{figure}
The spectral linewidth versus total bias current is presented in Fig. \ref{df2}. Similarly to the previous case, the linewidth for the uniform current profile is the smallest one. The increase of the linewidth in the nonuniform case in comparison to the linewidth of a uniform profile (in the working range of bias current $\eta_0 \sim 0.15 \div 0.22$) is about two times if the electrode is half of the junction $x_{2} - x_{1}= L/2$ (Fig. \ref{df2}, dashed curve), and about three times if $x_{2} - x_{1} < L/4$ in the working range of bias current $\eta_0 \sim 0.13 \div 0.17$  (Fig. \ref{df2}, short-dashed curve, dotted curve). Thus, to obtain a broad spectral line the width of the electrode must be significantly smaller than the junction length.

The power of the FFO radiation is shown in the inset of Fig. \ref{df2}. As can be seen from the figure, the maximum output power for narrow electrodes $x_{2}-x_{1} <L/4$ is comparable with the radiation power for wide electrode $x_{2} - x_{1}= L/2$ and roughly two times smaller than the maximum power for uniform current distribution case.

\begin{figure}[ht]
\resizebox{1\columnwidth}{!}{
\includegraphics{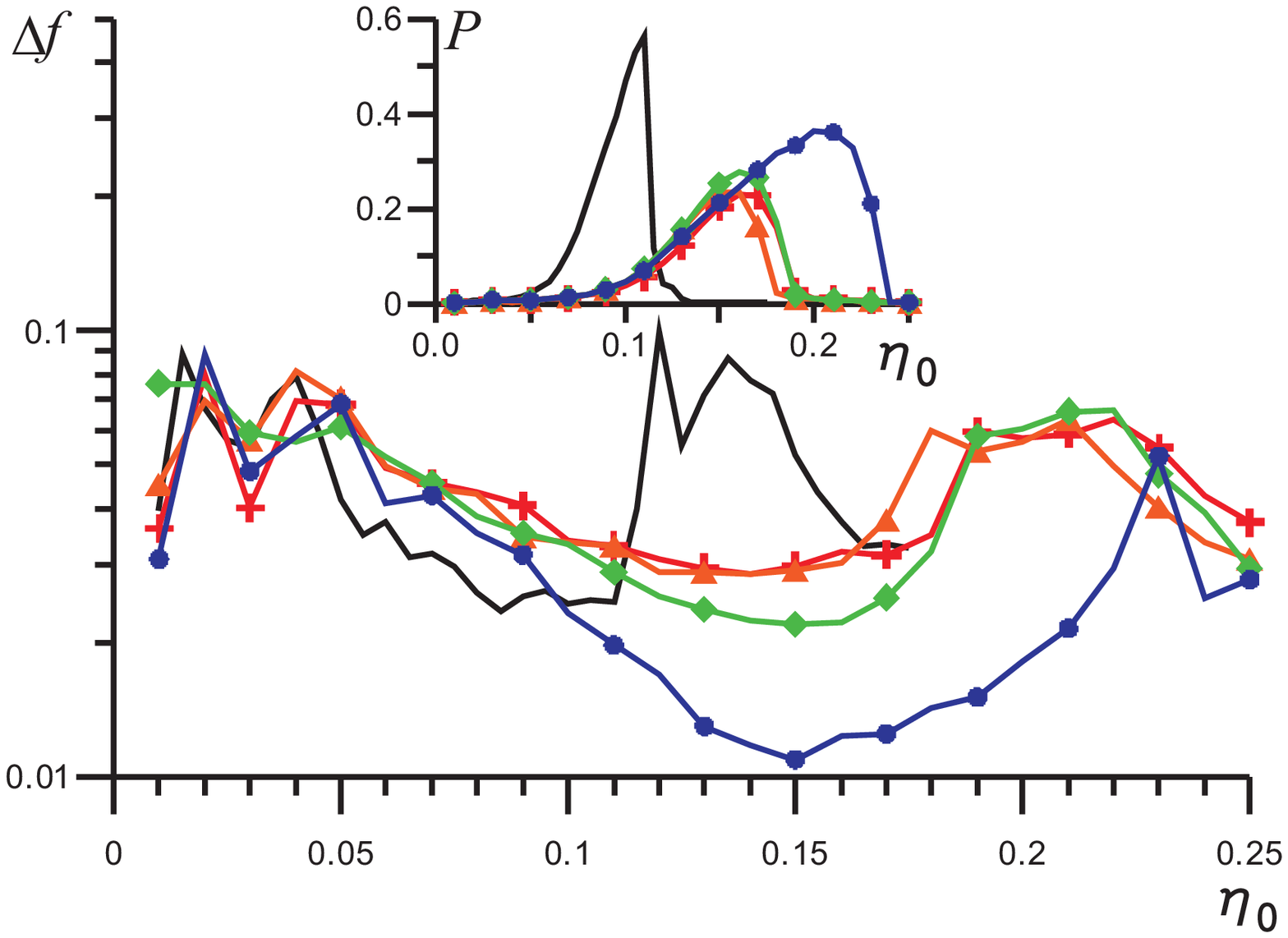}}
{\caption{FFO linewidth for the extremely narrow electrode $x_{1}=x_{2}=40$ and different decay factors $p$: curve with circles - $p = 0.03$, curve with diamonds - $p = 0.1$, curve with triangles - $p = 0.3$, curve with crosses - $p = 3$. The linewidth for the inline JTJ is presented for comparison. Inset: Radiated power with the same notations.}
\label{decay}}
\end{figure}
The last case studied in this paper is the dependence of the spectral and power characteristics on the decay factor $p$ of the current profile. In Fig. \ref{decay} the FFO linewidth for $x_{1}=x_{2}=40$ and different laws of current decay is presented by curves with symbols. As can be seen, for smaller $p$ factor the obtained spectral line is more narrow, while the radiated power is larger (see the inset of Fig. \ref{decay}). For the factor $p \ge 0.3$ the linewidth ceases to increase and the power almost ceases to fall. In the limit of large $p$ the bias current profile turns into a short peak rapidly decreasing away from the junction edge. In this case it is interesting to compare an overlap JTJ with the extremely narrow current electrode shifted to the input (right) edge with the JTJ of the inline geometry. In the inline case the bias current is injected into the junction parallel to its long direction, which in one-dimensional model leads to injection at the ends of the junction only, so the bias current profile is described by delta functions at the ends of the junction \cite{types}: $\eta(x)=\eta_0L[\delta(x)+\delta(x-L)]$. It has been shown \cite{inline} that for the inline JTJ the linewidth is 2-2.5 times larger than for the overlap case of uniform distribution with a small difference in the radiated power. In Fig. \ref{decay} the case of the inline geometry is presented by thin solid curve for comparison with the optimized cases of overlap JTJ. It is seen that the linewidth for nonuniform bias current distribution with rapid current decay (curves with crosses and triangles) is slightly larger than the inline FFO linewidth, while the inline geometry gives significantly larger power (presumably due to the fact that for the inline case the fluxons are accelerated from both junction ends).

In conclusions, the possibility of increasing of the FFO linewidth at the quasimonochromatic generation regime by supplying a nonuniform bias current profile has been studied. For this purpose, the output power and spectral characteristics for different current profiles have been computed and compared. It is shown that the optimization of the nonuniform current profile allows increasing the linewidth by a factor of three with the moderate power reduction, which opens up new possibilities for solving the task of creation of noisy non-stationary spectrometer based on a long Josephson junction. On the other hand, for a number of applications a three times increase of the linewidth (which in practice leads to the maximal linewdith of order 100-200 MHz in 450-700 GHz frequency range for Nb-AlOx-Nb FFOs) may be not enough and here either additional mixing with low frequency noise or transfer to high temperature superconductors can help.

The work was supported by RFBR (Project No. 09-02-00491), Human Capital Foundation, Dynasty Foundation and by the Act 220 of Russian Government (project 25).

\end{document}